\documentclass[preprint,amsmath,amssymb]{revtex4}
\usepackage{graphicx}
\usepackage{dcolumn}
\usepackage{bm}

\newcommand{\vek}[1]{\mbox{\bf{#1}}}
\def\KK{\textsc{kk}}
\def\kkonium{\textsc{kkonium}}
\def\b{\textsc{b}}

\begin{document}

\title{Kaluza-Klein Burst: a New Mechanism \\ for Generating Ultrahigh-Energy
Cosmic Rays}

\author{Je-An Gu} %
\email{jagu@phys.ntu.edu.tw} %
\affiliation{Department of Physics, National Taiwan University,
Taipei 106, Taiwan, R.O.C.}

\date{\today}

\begin{abstract}
By invoking small extra dimensions as a good energy bearer, a new
scenario for understanding the origin of ultrahigh-energy cosmic
rays (UHECRs), from both the bottom-up and the top-down
viewpoints, is proposed. We explore the possibility of generating
UHECRs via Kaluza-Klein (KK) bursts, a violent energy transfer
from extra dimensions to ordinary dimensions through collisions
between KK modes, in particular, within clumps of KK modes. The
possible scales of these clumps range from the astronomical, e.g.\
KK stellar compact objects, to the microscopic, e.g.\
``\kkonium''. Advantages of this KK burst model of UHECRs and
possible signatures of clumped KK modes are discussed.
\end{abstract}

\pacs{98.70.Sa}

\maketitle


\section{Introduction} \label{Introduction}

More than one thousand ultrahigh-energy cosmic ray (UHECR) events
of energies above $10^{19}\,$eV have been detected in the past ten
years. In particular the events above $10^{20}\,$eV
\cite{Takeda:1998ps,Abbasi:2002ta} have fired the controversy
about the existence and the origin of the Greisen-Zatsepin-Kuzmin
(GZK) cutoff \cite{GZK}. No matter whether the cutoff exists or
not, the detection of UHECRs has brought various challenges and
inspired a lot of ideas and imaginations for an ultrahigh-energy
part of the world. (For a review, see \cite{Anchordoqui:2002hs}.)

The difficulties in UHECR models mainly stem from two
observational results, ultrahigh energies and roughly isotropic
arrival directions. Ultrahigh energies make it difficult, in the
bottom-up models (for a review, see
\cite{Bhattacharjee:1998qc,Olinto:2000sa}), to find powerful
enough sources, in particular, under continuous energy losses via
the GZK mechanism. This difficulty leads to the GZK puzzle. In
addition, isotropic arrival directions make it unlikely to explain
the UHECR events via only a few sources, unless their trajectories
can have been significantly deflected by cosmic magnetic fields.

The situation is different in the top-down models (for a review,
see \cite{Bhattacharjee:1998qc}), where UHECRs are produced via
decays or annihilation of very massive particles (or topological
defects) so that energy is not an issue. These very massive
particles might behave like dark matter and reside in the local
dark halo. In this case the roughly isotropic distribution of
arrival directions is not an issue, either. Nevertheless, we know
little yet about such exotic, very massive particles. Their
existence is unconfirmed.

In this paper we propose a new mechanism for generating UHECRs by
invoking small compact extra (spatial) dimensions %
\footnote{The existence of extra dimensions is suggested by
various theories beyond the Standard Model of particle physics
such as string theory.} %
as a good energy bearer. %
We explore the feasibility of generating UHECRs via Kaluza-Klein
(KK) bursts, a violent energy transfer from extra dimensions to
ordinary dimensions through collisions of KK modes. We will show
how KK bursts can serve and benefit the construction of UHECR
models, including both the bottom-up and the top-down models,
without introducing new particle species.

The paper is organized as follows. Sec.\ \ref{KKBurst} sketches
the basic idea of the KK burst model of UHECRs. Sec.\
\ref{Spectrum} contains the study of the spectrum of UHECRs
originated from KK bursts involving uniformly distributed KK
modes, clumped KK modes, and KK compact objects (in particular,
``\kkonium''), respectively. A summary and discussions follow in
Sec.\ \ref{summary}.

\section{Kaluza-Klein Burst}
\label{KKBurst}

As a quantum nature, an excitation in an extra dimension of a
finite size $l_{\textsc{ED}}$, i.e.\ a Kaluza-Klein (KK) mode,
carries quantized KK momentum: $P_{\KK} = 2 \pi n_{\KK} /
l_{\textsc{ed}} \equiv n_{\KK} \hat{P}_{\KK}$, where the KK
(quantum) number $n_{\KK} = 0,\pm 1, \pm 2,\ldots\,$, and
$\hat{P}_{\KK}$ denotes one unit of KK momentum. We will use the
symbol KK$^{n}$ to denote a KK mode with a KK number $n$. KK$^{n}$
and KK$^{-n}$ will be called KK mode and anti-KK mode. %
(Note that KK modes and anti-KK modes need not to be particles and
anti-particles.)
A bound state consisting of a KK mode and an anti-KK mode will be
called ``\kkonium''.

As required by momentum conservation in extra dimensions, the KK
number is conserved module two, i.e., KK modes must be created or
annihilated in pairs. This feature helps extra dimensions be a
good energy bearer against the energy pillage by the GZK mechanism
and other similar energy degrading processes (as to be called
GZK-like mechanisms). In a usual way of thinking about the
GZK-like mechanisms, ultrahigh-energy particles lose energy
through the scattering with background particles or fields.
Nevertheless, due to KK momentum conservation, the energy stored
in extra dimensions is hardly dissipated through the interactions
with background particles that are mostly KK zero modes.

This feature can help the bottom-up scenario get rid of two
essential difficulties. In the bottom-up models, particles are
accelerated to ultrahigh energies within extreme astrophysical
environments, which are usually very dense. How ultrahigh-energy
particles escape from these dense regions without losing much
energy through the scattering with particles therein is a serious
intrinsic problem. Another difficulty stems from the GZK mechanism
that makes it unlikely for UHECRs to maintain energies beyond the
GZK threshold (around $10^{20}\,$eV) after travelling a distance
longer than 50 Mpc. Unfortunately, there are very few powerful
enough sources within the GZK zone, a region with a radius of 50
Mpc around the earth. These few sources can hardly explain the
UHECR spectrum and the distribution of arrival directions.
Nevertheless, if these ultrahigh energies are stored in KK
momentum of KK modes, they will have much better chance to be
carried out of the sources and to travel a long distance even much
larger than 50 Mpc without significant loss.

In addition, a KK (nonzero) mode might collide with an anti-KK
mode, through which the energy stored in extra dimensions may be
transferred to ordinary dimensions. If the size of extra
dimensions is extremely small such that $\hat{P}_{\KK}$ is
extremely large correspondingly, this energy transfer will be very
violent. For a three-dimensional observer, this process looks like
a burst caused by a collision between two very heavy particles
(with large effective mass from KK momentum), through which a
large amount of energy is released from the mass of incoming
particles to the kinetic energy of outgoing particles. This burst
is dubbed ``\textsf{Kaluza-Klein (KK) burst}'' hereafter. This
violent energy transfer from extra dimensions to ordinary
dimensions provides a mechanism for generating UHECRs.

%

KK modes with ultra-large KK momentum may be created, in the form
of free KK modes or \kkonium, from the very early universe or
within extreme astrophysical environments (which may locate
outside the GZK zone). The UHECRs detected on the earth are those
generated via KK bursts involving KK modes which have come into or
are always within the GZK zone.

KK bursts serve both the top-down and the bottom-up models. %
In the top-down point of view, the process of generating UHECRs
from the KK burst of two freely moving KK modes is similar to that
involving annihilation of superheavy particles
\cite{Blasi:2001hr}, while the case of KK bursts in \kkonium,
i.e.\ decays of \kkonium, is similar to decays of superheavy
particles \cite{Berezinsky:1997hy,Kuzmin:1997cm}.

For the bottom-up scenario, the detected UHECRs are generated via
KK bursts within the GZK zone, involving KK modes or \kkonium\
which are created in extreme astrophysical environments, whose
locations may be outside the GZK zone. %
Note that in this model the extreme astrophysical environments are
sources of KK modes and \kkonium\ that generate UHECRs within the
GZK zone, but not the ``first-hand'' sources of UHECRs. %
Also note that in the KK burst model UHECRs are KK zero modes, but
not KK (nonzero) modes. Thus, in this model UHECRs do still suffer
the energy pillage by the GZK-like mechanisms. It is the KK
(nonzero) modes, the source of UHECRs, but not UHECRs themselves,
that can defend the energy they carry against the GZK-like
mechanisms during the journey to the earth.

The way of circumventing the GZK-like energy pillage via small
extra dimensions in the KK burst model is similar to that in the
Z-burst model of UHECRs
\cite{Weiler:1982qy,Fargion:1997ft,Weiler:1997sh}. The Z-burst
model explains the detection of the super-GZK events by invoking
(extremely-high-energy) neutrinos as a good energy messenger,
which, in the KK burst model, is played by KK modes (including
\kkonium).

%

\section{Spectrum} \label{Spectrum}

In calculating the differential flux of UHECRs originated from KK
bursts, we consider the simplest channel: a KK$^{+1}$ mode
collides with a KK$^{-1}$ mode, creating two outgoing KK zero
modes with large kinetic energy, which then produce two leading
jets through fragmentation. We assume that the distributions of
KK$^{\pm 1}$ modes are both described by the number density
$n_{\KK} (\vek{r})$, where $\vek{r}$ is the position vector
measured by observers on the earth. In addition, we assume that KK
momentum dominates the energy of each KK$^{(\pm 1)}$ mode, such
that the gain of large kinetic energy for each outgoing KK zero
mode mainly comes from KK momentum.

The process of a KK burst is similar to pair annihilation of two
heavy particles with (effective) mass about $\hat{P}_{\KK}$. (Note
that, unlike pair annihilation, KK bursts may involve different
species of particles, e.g., an electron and a photon.) The
possibility that UHECRs originate from the annihilation of
superheavy dark matter has been considered by Blasi, Dick, and
Kolb \cite{Blasi:2001hr}. It is straightforward to employ the
formulation therein to study the spectrum of UHECRs generated by
KK bursts. Following \cite{Blasi:2001hr}, the formula for the
spectrum of UHECRs originated from KK bursts can be written as
\begin{eqnarray} \label{spectrum}
\mathcal{F} &=&  \frac{d \mathcal{N} \left(
E,E_{\textrm{jet}}=\hat{P}_{\KK} \right)
}{dE} \cdot F \nonumber \\
&=& \frac{d \mathcal{N} \left( E,E_{\textrm{jet}}=\hat{P}_{\KK}
\right) }{dE} \cdot 2 \, \langle \sigma_{\b} v \rangle \int d^3
\vek{r} \frac{n_{\KK}^2 (\vek{r})}{4 \pi \left| \vek{r} \right|
^2} \; ,
\end{eqnarray}
where $d \mathcal{N} (E,E_{\textrm{jet}})/dE$ is the fragmentation
spectrum from a jet of energy $E_{\textrm{jet}}$
\cite{Dokshitzer:wu} and $F$ denotes the UHECR flux. As indicated
in the above equation, for calculating the spectrum $\mathcal{F}$
we need to know the KK burst cross-section $\sigma_{\b}$ and the
number density of KK$^{\pm 1}$ modes $n_{\KK}$. So far the
knowledge about the cross-section at such high energies is poor.
In the following calculations, we will introduce the unitary
bound, $\hat{P}_{\KK}^{-2}$, to the KK burst cross-section and
utilize the ratio $\alpha_{\KK} \equiv \langle \sigma_{\b} v
\rangle / \hat{P}_{\KK}^{-2}$, which can be regarded as an
effective coupling constant associated with KK bursts. As to
$n_{\KK}$, we will introduce the energy density of dark matter to
give an upper limit to the energy density of KK modes $\rho_{\KK}$
(i.e.\ $n_{\KK}\hat{P}_{\KK}$).

As indicated in \cite{Blasi:2001hr}, for the consistency with the
observed UHECR spectrum in the model of superheavy dark matter
annihilation, the preferred value of the mass of superheavy dark
matter is in the range $10^{21}$--$10^{23}\,$eV, and the
corresponding value of the flux $F$ ranges from $0.4$
$\textrm{km}^{-2} \textrm{yr}^{-1}$ (for mass $10^{21}\,$eV) to
$7$ $\textrm{km}^{-2} \textrm{yr}^{-1}$ (for mass $10^{23}\,$eV).
These results can be applied to the KK burst model. Accordingly,
the preferred value of one-unit KK momentum $\hat{P}_{\KK}$ is in
the range $10^{21}$--$10^{23}\,$eV, i.e.\ the corresponding size
of extra dimensions ranges from $10^{-27}\,$cm to $10^{-25}\,$cm,
and the flux $F$ provided by KK bursts needs to reach the value
mentioned above, i.e.,
\begin{equation} \label{required-F}
\begin{array}{rcl}
\hspace{-0.7em} F &=& 0.4 \textrm{ km}^{-2} \textrm{yr}^{-1} \;
\mbox{ for } \; \hat{P}_{\KK} = 10^{21} \mbox{eV} \, %
\mbox{
$\sim 10^{25}$cm$^{-1}$} %
\, , \\
\hspace{-0.7em} F &=& \hspace{0.33em} 7 \hspace{0.44em} \textrm{
km}^{-2} \textrm{yr}^{-1} \; \mbox{ for } \; \hat{P}_{\KK} =
10^{23} \mbox{eV} \, %
\mbox{
$\sim 10^{27}$cm$^{-1}$} \, .
\end{array}
\end{equation}

\subsection{Uniformly Distributed KK Modes}

In the following we study two parts of contributions of KK bursts
to UHECRs, one from KK modes in the local dark halo and the other
from those in the GZK zone (excluding the effect of the
overdensity in the local halo). We first consider a toy model
where KK$^{\pm 1}$ modes are uniformly distributed with a constant
number density $n_{\KK \textrm{(halo)}}$ in the local halo. In
this model, the flux $F$ is simply
\begin{equation} \label{toy-F}
F_{\textrm{halo}} = 2   \, \langle \sigma_{\b} v \rangle \, n_{\KK
\textrm{(halo)}}^2 \int_{\textrm{halo}} \frac{d^3 \vek{r}}{4 \pi
\left| \vek{r} \right| ^2} \; .
\end{equation}
Choosing the characteristic size of the local halo to be 30 kpc
and the distance of the solar system from the halo center to be 8
kpc, we obtain
\begin{equation} \label{toy-F-value in halo}
F_{\textrm{halo}} \simeq \left( \frac{6 \times
10^{-26}}{\textrm{km}^{2} \, \textrm{yr}} \right) \cdot \,
\alpha_{\KK} \; \xi_{\KK \textrm{(halo)}}^2 \cdot \left(
\frac{\hat{P}_{\KK}}{10^{21} \mbox{eV}} \right) ^{-4} ,
\end{equation}
where $\alpha_{\KK} \equiv \langle \sigma_{\b} v \rangle /
\hat{P}_{\KK}^{-2}$ and $\xi_{\KK \textrm{(halo)}} \equiv
\rho_{\KK \textrm{(hlao)}} / \rho_{\textrm{halo}}$ (i.e.\ $n_{\KK
\textrm{(halo)}} \hat{P}_{\KK} / \rho_{\textrm{halo}}$), which is
the fraction of the energy density contributed from KK modes in
the local halo. Note that in the above equation we have introduced
the unitary bound $\hat{P}_{\KK}^{-2}$ to the KK burst
cross-section $\langle \sigma_{\b} v \rangle$ and the energy
density of the local halo,
$\rho_{\textrm{halo}}=2\mbox{--}13\times 10^{-25}$ g cm$^{-3}$, as
an upper limit to the energy density of KK modes in the halo
$\rho_{\KK \textrm{(halo)}}$. As indicated in Eq.\
(\ref{toy-F-value in halo}), the flux $F$ is proportional to
$\hat{P}_{\KK}^{-4}$ for fixed $\alpha_{\KK}$ and $\xi_{\KK
\textrm{(halo)}}$. The value $6 \times 10^{-26} \mbox{ km$^{-2}$}
\mbox{ yr$^{-1}$}$, which can be regarded as an upper bound of the
flux $F$ for $\hat{P}_{\KK} \geq 10^{21} \, \mbox{eV}$, is far
below the required values in Eq.\ (\ref{required-F}).

In the same way, we can calculate the flux contributed from the
uniformly distributed KK modes within the GZK zone (ignoring the
overdensity in the local halo). We obtain
\begin{equation} \label{toy-F-value in GZK}
F_{\textsc{gzk}} \simeq \left( \frac{7 \times
10^{-33}}{\textrm{km}^{2} \, \textrm{yr}} \right) \cdot \,
\alpha_{\KK} \; \xi_{\KK \textsc{(gzk)}}^2 \cdot \left(
\frac{\hat{P}_{\KK}}{10^{21} \mbox{eV}} \right) ^{-4} ,
\end{equation}
where $\xi_{\KK \textsc{(gzk)}} \equiv \rho_{\KK \textsc{(gzk)}} /
\rho_{\textsc{dm}}$, which is the fraction of the energy density
of dark matter contributed from KK modes within the GZK zone. Here
we have introduced the energy density of the dark matter in the
universe, $\rho_{\textsc{dm}}=1.4\mbox{--}4.3\times 10^{-30}$ g
cm$^{-3}$, as an upper limit to the energy density of KK modes
within the GZK zone, $\rho_{\KK \textsc{(gzk)}}$. This flux is in
general much smaller than that originated from the local halo, and
is also far below the required values in Eq.\ (\ref{required-F}).

\subsection{Clumped KK modes and KK Compact Objects}

KK modes may clump together in small regions and form compact
objects through gravity or other interactions. The large
overdensity in these clumps can enhance the flux $F$ because of
the non-linearity with respect to $n_{\KK}$ in Eq.\
(\ref{spectrum}).

We first consider the sub-galactic clumps of KK modes in the local
halo. For simplicity we make the following assumptions: (i) KK
modes are uniformly distributed with a constant number density
$n'_{\KK}$ in each clump. (ii) Clumps are uniformly distributed in
the halo. (iii) All clumps possess the same (effective) mass. In
this case, the formula for the flux $F$ is:
\begin{eqnarray} \label{clump-F-halo}
F_{\textrm{halo}} &=& \left( \frac{n'_{\KK}}{n_{\KK}}
\right)_{\textrm{halo}} \cdot 2 \, \langle \sigma_{\b} v \rangle
\, n_{\KK \textrm{(halo)}}^2 \int_{\textrm{halo}} \frac{d^3
\vek{r}}{4 \pi \left| \vek{r} \right| ^2} \nonumber \\
&\simeq& \left( \frac{n'_{\KK}}{n_{\KK}} \right)_{\textrm{halo}}
\cdot \left( \frac{6 \times 10^{-26}}{\textrm{km}^2 \,
\textrm{yr}} \right) \, \cdot \, \alpha_{\KK} \, \xi_{\KK
\textrm{(halo)}}^2
\, \left( \frac{\hat{P}_{\KK}}{10^{21} \textrm{eV}} \right) ^{-4} . %
\end{eqnarray}
Compared with Eqs.\ (\ref{toy-F}) and (\ref{toy-F-value in halo}),
the above formula contains an extra factor $(n'_{\KK} / n_{\KK}
)_{\textrm{halo}}$ that is the ratio of the
overdensity $n'_{\KK}$ in each clump to the mean 
density $n_{\KK \textrm{(halo)}}$ in the halo. This factor can
enhance the flux substantially if the sub-clumps of KK modes are
over-dense significantly.

The required value of the flux $F$ in Eq.\ (\ref{required-F})
gives a lower bound to the ratio $(n'_{\KK} / n_{\KK}
)_{\textrm{halo}}$ and the corresponding energy density of KK
modes in each clump in the halo, $(\rho'_{\KK})_{\textrm{halo}}$
[i.e.\ $(n'_{\KK} \hat{P}_{\KK})_{\textrm{halo}}$], as follows:
\begin{subequations} \label{required n ratio-halo}
\begin{eqnarray}
\left( n'_{\KK}/n_{\KK}
\right)_{\textrm{halo}}^{(10^{21}\textrm{eV})} &\simeq& %
7 \times 10^{24} \cdot \, \alpha_{\KK}^{-1} \; \xi_{\KK
\textrm{(halo)}}^{-2} \, \geq \, 7 \times 10^{24}
\, , %
\label{required n ratio-halo 10^21 eV} \\
\left( n'_{\KK}/n_{\KK}
\right)_{\textrm{halo}}^{(10^{23}\textrm{eV})} &\simeq& %
\hspace{1em} 10^{34} \hspace{0.8em} \cdot \, \alpha_{\KK}^{-1} \;
\xi_{\KK \textrm{(halo)}}^{-2} \, \geq \, \hspace{0.4em} 10^{34}
\hspace{0.3em} \hspace{1em} \, , %
\label{required n ratio-halo 10^23 eV}
\end{eqnarray}
\end{subequations}
or, equivalently, in unit of \ g cm$^{-3}$,
\begin{subequations} \label{required rhoKK - halo}
\begin{eqnarray}
\hspace{-2em} %
\left( \rho'_{\KK} \right)_{\textrm{halo}}^{(10^{21}\textrm{eV})}
&\simeq& \hspace{1.31em} 4 \hspace{1.31em} \cdot \,
\alpha_{\KK}^{-1} \; \xi_{\KK \textrm{(halo)}}^{-1} \, \geq \,
\hspace{0.2em} 4 \, ,
\label{required rhoKK - halo 10^21 eV} \\
\hspace{-2em} %
\left( \rho'_{\KK} \right)_{\textrm{halo}}^{(10^{23}\textrm{eV})}
&\simeq& 6 \times 10^{9} \cdot \, \alpha_{\KK}^{-1} \; \xi_{\KK
\textrm{(halo)}}^{-1} \, \geq \, 6 \times 10^{9} \, , %
\label{required rhoKK - halo 10^23 eV}
\end{eqnarray}
\end{subequations}
where the superscripts denote the conditions for $\hat{P}_{\KK}
\,$.

In the same way, the contribution from the clumps of KK modes in
the GZK zone (excluding the effect of the overdensity in the local
halo) is:
\begin{equation} \label{clump-F-GZK}
F_{\textsc{gzk}} \simeq \left( \frac{n'_{\KK}}{n_{\KK}}
\right)_{\textsc{gzk}} \cdot \left( \frac{7 \times
10^{-33}}{\textrm{km}^2 \, \textrm{yr}} \right) \, \cdot \,
\alpha_{\KK} \, \xi_{\KK \textsc{(gzk)}}^2
\, \left( \frac{\hat{P}_{\KK}}{10^{21} \textrm{eV}} \right) ^{-4} , %
\end{equation}
where $(n'_{\KK}/n_{\KK})_{\textsc{gzk}}$ is the ratio of the
overdensity in each clump to the mean density in the GZK zone. The
requirement in Eq.\ (\ref{required-F}) gives a lower bound to the
ratio $(n'_{\KK} / n_{\KK} )_{\textsc{gzk}}$ and the corresponding
energy density of KK modes in each clump in the GZK zone,
$(\rho'_{\KK})_{\textsc{gzk}}$ [i.e.\ $(n'_{\KK}
\hat{P}_{\KK})_{\textsc{gzk}}$], as follows:
\begin{subequations} \label{required n ratio-GZK}
\begin{eqnarray}
\left( n'_{\KK}/n_{\KK}
\right)_{\textsc{gzk}}^{(10^{21}\textrm{eV})} &\simeq& %
6 \times 10^{31} \cdot \, \alpha_{\KK}^{-1} \; \xi_{\KK
\textsc{(gzk)}}^{-2} \, \geq \, 6 \times 10^{31} \, , %
\label{required n ratio-GZK 10^21 eV} \\
\left( n'_{\KK}/n_{\KK}
\right)_{\textsc{gzk}}^{(10^{23}\textrm{eV})} &\simeq& %
\; \; \; 10^{41} \hspace{0.9em} \cdot \, \alpha_{\KK}^{-1} \;
\xi_{\KK \textsc{(gzk)}}^{-2} \, \geq \, \hspace{0.2em} 10^{41}
\hspace{1.45em} \, , %
\label{required n ratio-GZK 10^23 eV}
\end{eqnarray}
\end{subequations}
or, equivalently, in unit of \ g cm$^{-3}$,
\begin{subequations} \label{required rhoKK - GZK}
\begin{eqnarray}
\hspace{-2em} %
\left( \rho'_{\KK} \right)_{\textsc{gzk}}^{(10^{21}\textrm{eV})}
&\simeq& \hspace{1em} 200 \hspace{1em} \cdot \, \alpha_{\KK}^{-1}
\; \xi_{\KK \textsc{(gzk)}}^{-1} \, \geq \, \hspace{0.2em} 200 \, , %
\label{required rhoKK - GZK 10^21 eV} \\
\hspace{-2em} %
\left( \rho'_{\KK} \right)_{\textsc{gzk}}^{(10^{23}\textrm{eV})}
&\simeq& 3 \times 10^{11} \cdot \, \alpha_{\KK}^{-1} \; \xi_{\KK
\textsc{(gzk)}}^{-1} \, \geq \, 3 \times 10^{11} \, . %
\label{required rhoKK - GZK 10^23 eV}
\end{eqnarray}
\end{subequations}

The number density ratios in Eqs.\ (\ref{required n ratio-halo})
and (\ref{required n ratio-GZK}) are not so large as they look
like. For comparison, we list in the following table the minimal
values of the above ratios as well as the nucleon number density
ratios of several stellar objects (the sun, the earth, white
dwarfs, and neutron stars) to the background (the galactic disk
and the universe).
\begin{center}
\begin{tabular}{|ccc|} \hline
\hspace*{3.5em} \footnotesize{in unit of:} %
& $n_{\KK}$\footnotesize{(halo)} &
$n_{\KK}$\footnotesize{(\textsc{gzk})}
\\ \hline
$n'_{\KK}\scriptstyle{(\hat{P}_{\KK}=10^{21}\textrm{eV})}$
                        & $7 \times 10^{24}$
                        & $6 \times 10^{31}$
\\ 
$n'_{\KK}\scriptstyle{(\hat{P}_{\KK}=10^{23}\textrm{eV})}$
                        & $10^{34}$
                        & $10^{41}$
\\ \hline \hline
\hspace*{3.5em} \footnotesize{in unit of:} %
            & $n_{\textrm{N}}$\footnotesize{(disk)}
            & $n_{\textrm{B}}$\footnotesize(universe)
\\ \hline

$n_{\textrm{N}}$\footnotesize{(sun)} \ \ &  1--4$\times 10^{23}$
                                         &  3--8$\times 10^{30}$
\\ 
$n_{\textrm{N}}$\footnotesize{(earth)} & 0.4--2$\times 10^{24}$
                                       & 1--3$\times 10^{31}$
\\ 
$n_{\textrm{N}}$\footnotesize{(white dwarf)}
                    & $10^{24.5}$--$10^{32}$
                    & $10^{32}$--$10^{39}$
\\ 
$n_{\textrm{N}}$\footnotesize{(neutron star)}
                    & $10^{33}$--$10^{38}$
                    & $10^{40}$--$ 10^{45}$
\\ \hline
\end{tabular}
\\ \vspace{0.5ex}
\begin{tabular}{rcl}
$n_{\textrm{N}}$\footnotesize{(disk)} & = & $2\mbox{--}7$ cm$^{-3}$ \\
$n_{\textrm{B}}$\footnotesize{(universe)} & = & $1.1\mbox{--}2.6
\times 10^{-7}$ cm$^{-3}$
\end{tabular}
\end{center}
Here $n_{\textrm{N}}$\footnotesize{(disk)} \normalsize is the
nucleon number density of the galactic disk, and $n_{\textrm{B}}$
is the baryon number density of the universe. We can see that, in
both the case of the local galaxy and the case of the GZK zone,
the minimal value of the ratio $n'_{\KK}$/$n_{\KK}$ is comparable
to the nucleon number density ratio of the earth or a low-density
white dwarf to the background (the galactic disk or the universe)
for $\hat{P}_{\KK}=10^{21}\,$eV, and is comparable to that of a
low-density neutron star for $\hat{P}_{\KK}=10^{23}\,$eV. This is
a hint that the detected UHECRs are generated via KK bursts within
some sorts of stellar compact objects consisting of KK modes:
``\textsc{kk star}s'', ``\textsc{kk dwarf}s'', ``\textsc{kk
macho}s'' (massive compact halo objects made of KK modes), etc.

This hint is also provided by comparing energy densities in the
following table.
\begin{center}
\begin{tabular}{|ccccc|}
\multicolumn{3}{l}{(in unit of \ g cm$^{-3}$)}%
\\ \hline
  & halo\scriptsize{(10$^{21}$eV)} & \textsc{gzk}\scriptsize{(10$^{21}$eV)}
  & halo\scriptsize{(10$^{23}$eV)} & \textsc{gzk}\scriptsize{(10$^{23}$eV)}
\\ \hline
  $\rho'_{\KK}$ %
    & 4 & 200 & $6 \times 10^{9}$ & $3 \times 10^{11}$ %
\\ \hline \hline
    & sun & earth & white dwarf & neutron star
\\ \hline
  $\rho_{\textrm{mean}}$ %
    & 1.4 & 5.5 & 30--10$^{9}$ & 10$^{10}$--10$^{15}$
\\ \hline
\end{tabular}
\end{center}
This table indicates that for $\hat{P}_{\KK}=10^{21}\,$eV the
minimal energy density of clumped KK modes is comparable to the
mass density of the earth or a low-density white dwarf, and for
$\hat{P}_{\KK}=10^{23}\,$eV it increases to the value comparable
to that of a high-density white dwarf or a low-density neutron
star.

In addition to KK stellar compact objects, the production of
UHECRs via KK bursts of clumped KK modes may also occur within
much smaller compact objects, even as tiny as \kkonium.

\subsection{KKONIUM}

During a KK burst within \kkonium, two originally bound KK modes
get a large amount of kinetic energy from extra dimensions and
become two unbound, extremely energetic KK zero modes. For a
three-dimensional observer, this process looks like the decay of a
superheavy particle (of mass about $2\hat{P}_{\KK}$ for the
lightest \kkonium). The possibility that UHECRs are produced at
the decays of metastable superheavy particles has been proposed by
Berezinsky, Kachelrie\ss\ and Vilenkin \cite{Berezinsky:1997hy}
and by Kuzmin and Rubakov \cite{Kuzmin:1997cm}. The decay spectrum
was calculated by Birkel and Sarkar \cite{Birkel:1998nx} using
HERWIG program. (For further calculations using DGLAP evolution
equations, see \cite{DGLAP}.) If these superheavy particles
account for the dark matter in the local halo and their decays are
responsible for the detected UHECRs, their mass $M_{X}$ should be
within $10^{21}\,$--$10^{23}\,$eV and the corresponding decay
width $\Gamma_{X}$ is:
\begin{subequations}
\begin{eqnarray}
\Gamma_{X}^{-1} &=& \hspace{0.9em} 10^{20} \hspace{0.86em} \mbox{
yr } \quad \mbox{for} \quad M_{X} = 10^{21} \, \mbox{eV} \, , \\
\Gamma_{X}^{-1} &=& 8 \times 10^{20} \mbox{ yr } \quad \mbox{for}
\quad M_{X} = 10^{23} \, \mbox{eV} \, .
\end{eqnarray}
\end{subequations}

From the above information and the relation between the decay
width $\Gamma$, the mass $M$ and the characteristic size $R$ of a
\kkonium,
\begin{equation}
\Gamma \sim M^{-2} R^{-3} \, ,
\end{equation}
we can estimate the requisite value of $R$ for the case where the
decays of \kkonium\ account for the generation of UHECRs. For
\kkonium\ systems consisting of KK$^{\pm 1}$ modes, we obtain
\begin{subequations}
\begin{eqnarray}
R \sim 10^{-5} \; \alpha_{\KK}^{2/3} \mbox{ cm} %
& \mbox{ for } & \hat{P}_{\KK} = 10^{21}/2 \, \mbox{ eV} \, , \\
R \sim 10^{-6} \; \alpha_{\KK}^{2/3} \mbox{ cm} %
& \mbox{ for } & \hat{P}_{\KK} = 10^{23}/2 \, \mbox{ eV} \, .
\end{eqnarray}
\end{subequations}
The above sizes are much larger than the Bohr radius of \kkonium,
which is characterized by the inverse of the KK momentum.
Consequently \kkonium\ as a UHECR source should be in a highly
excited state.

For the other case in which \kkonium\ systems contribute only a
tiny fraction of dark matter, the requisite decay width should be
much larger and accordingly the size of \kkonium\ should be much
smaller. This issue is of our great interest and is under
investigation.

\section{Discussion and Summary} \label{summary}

In this paper the feasibility of generating UHECRs via KK bursts
has been discussed. Basically we make use of extra dimensions to
overcome difficulties in UHECR models. The most essential
difficulty is the energy, its unusual largeness under the energy
pillage by the GZK-like mechanisms. Regarding the unusual
largeness of energy, the violent energy transfer from extremely
small extra dimensions to ordinary dimensions through KK bursts
can easily reach such high energies. Against the energy pillage,
extra dimensions, as benefiting from KK momentum conservation, can
protect the energy therein when KK (non-zero) modes travel through
the universe. This feature is of great benefit to the bottom-up
models because it can help particles of large energies (in the
form of KK momentum) escape from the source without losing much
energy and make it possible for UHECR sources (more precisely,
sources of KK modes and \kkonium\ that generate UHECRs within the
GZK zone) to be located outside the GZK zone. In this regard, KK
modes are messengers of ultrahigh energies for UHECRs generated
within the GZK zone, and accordingly they play a similar role to
the neutrinos in the Z-burst model
\cite{Weiler:1982qy,Fargion:1997ft,Weiler:1997sh}. In addition, KK
bursts might also be an origin of extremely-high-energy neutrinos
invoked in the Z-burst model. This is an interesting possibility
worthy of further studies.

In contrast to smoothly distributed KK modes that produce too few
UHECRs, clumped KK modes, of various sizes from KK stellar compact
objects to those as micro as \kkonium, are possible UHECR origins.
In this regard, KK bursts provide a possibility, without
introducing new particle species, of realizing both UHECR models
invoking annihilation of superheavy particles \cite{Blasi:2001hr}
and those invoking decays \cite{Berezinsky:1997hy,Kuzmin:1997cm}.
For the consistency with the observed spectrum, the minimal energy
density of KK stellar objects should range from the mass density
of a low-density white dwarf to that of a low-density neutron star
for the corresponding size of extra dimensions ranging from
$10^{-27}\,$cm to $10^{-25}\,$cm. In addition, \kkonium\ should be
in highly excited states with sizes nearly twenty orders of
magnitude larger than its Bohr radius. The detailed properties of
the \kkonium\ decay and the resultant UHECR spectrum are important
issues for a full understanding of \kkonium-originated UHECRs, and
therefore merit further investigations.

A homogeneous distribution of clumped KK modes within the GZK zone
can explain the isotropic arrival directions. In contrast, UHECRs
from clumped KK modes with an isotropic distribution in the galaxy
should be anisotropic: more from the side of the galactic center,
while less from the other side. Furthermore, UHECRs from KK
compact objects should exhibit the small-scale clustering in
arrival directions, which has been indicated by AGASA data
\cite{Hayashida:2000zr,Takeda:1999sg,Uchihori:1999gu}, and maybe
also the dispersion in arrival times of these clustering events.
These are possible characteristic signatures of KK compact
objects.

In this work we adopt QCD fragmentation to play a role in the
production of UHECRs after KK bursts. UHECRs originating from such
fragmentation are mostly photons and neutrinos, which seem to be
disfavored by the present data. Nevertheless, in the KK burst
scenario QCD fragmentation is not a necessary ingredient if the KK
modes involved are neither quarks nor gluons. It is possible that
KK modes can maintain their identities in particle species during
KK bursts, while only transferring momentum from extra dimensions
to ordinary dimensions. In this case, different species of KK
modes in KK bursts provide different composition of UHECRs. For
clarification, possible composition of KK modes and other
fragmentation processes are under investigation.

To sum up, the KK burst model of UHECRs invokes extra dimensions
as a good energy bearer against the energy pillage occurring
within powerful sources of ultrahigh-energy particles and in the
journey to the earth, thereby benefiting the bottom-up models. KK
bursts can generate extremely energetic particles that result in
observed UHECR spectrum, thereby making it possible to build a
top-down model without introducing new particle species. In
particular, the KK burst model simultaneously realizes the
following three scenarios for the origin of UHECRs:
\begin{enumerate}
\item[(a)] {\bf Decays of superheavy particles}
\cite{Berezinsky:1997hy,Kuzmin:1997cm}, which, in the KK
burst model, are played by \kkonium. %
\item[(b)] {\bf Annihilation of superheavy particles}
\cite{Blasi:2001hr}, which, in the KK burst model, are played by
KK modes and anti-KK modes. (Note that the particles involved in a
KK burst need not to be a pair of a particle and an anti-particle.) %
\item[(c)] {\bf Ultrahigh energies carried by messengers} (e.g.\
the Z-burst model
\cite{Weiler:1982qy,Fargion:1997ft,Weiler:1997sh}), which, in the
KK burst model, are played by KK modes and \kkonium.
\end{enumerate}
Thus, from both the bottom-up and the top-down viewpoints, the KK
burst model presents a new feasible way to understanding the
origin of UHECRs. On the other hand, UHECRs may encode information
about extra dimensions, and therefore might be a good probe of,
not only the ultrahigh-energy part, but also the extra-dimensional
part of the world.

\section*{Acknowledgments}
This work is supported in part by the CosPA project of the
Ministry of Education (MOE 89-N-FA01-1-4-3) and by the National
Science Council, Taiwan, R.O.C.\ (NSC 93-2811-M-002-056 and NSC
93-2112-M-002-047).

\end{document}